\documentclass[a4paper,12pt,useAMS]{emulateapj}
\usepackage{apjfonts,graphicx,graphics}

\usepackage{natbib}

\shorttitle{AMiBA Instrumentation}

\shortauthors{Chen et al.}

\begin{document}


\title{AMiBA: Broadband Heterodyne CMB Interferometry}


\author{ 
Ming-Tang Chen\altaffilmark{1},
Chao-Te Li\altaffilmark{1}, 
Yuh-Jing Hwang\altaffilmark{1},
Homin Jiang\altaffilmark{1},
Pablo Altamirano\altaffilmark{1}, 
Chia-Hao Chang\altaffilmark{1}, 
Shu-Hao Chang\altaffilmark{1}, 
Su-Wei Chang\altaffilmark{1}, 
Tzi-Dar Chiueh\altaffilmark{4}, 
Tah-Hsiung Chu\altaffilmark{4},
Chih-Chiang Han\altaffilmark{1}, 
Yau-De Huang\altaffilmark{1}, 
Michael Kesteven\altaffilmark{2}, 
Derek Kubo\altaffilmark{1}, 
Pierre Martin-Cocher\altaffilmark{1}, 
Peter Oshiro\altaffilmark{1}, 
Philippe Raffin\altaffilmark{1},
Tashun Wei\altaffilmark{1},
Huei Wang\altaffilmark{4}, 
Warwick Wilson\altaffilmark{2}, 
Paul T. P. Ho\altaffilmark{1,3}, 
Chih-Wei Huang\altaffilmark{4},
Patrick Koch\altaffilmark{1},
Yu-Wei Liao\altaffilmark{4}, 
Kai-Yang Lin\altaffilmark{1,4}, 
Guo-Chin Liu\altaffilmark{1}, 
Sandor M. Molnar\altaffilmark{1},
Hiroaki Nishioka\altaffilmark{1},
Keiichi Umetsu\altaffilmark{1}, 
Fu-Cheng Wang\altaffilmark{4},
Jiun-Huei Proty Wu\altaffilmark{4}
} 

\altaffiltext{1}{Academia Sinica, Institute of Astronomy and Astrophysics, P.O.Box 23-141, Taipei 106, Taiwan}

%%\altaffiltext{2}{Department of Physics, Institute of Astrophysics, \& Center for Theoretical Sciences, National Taiwan University, Taipei 10617, Taiwan} 

\altaffiltext{2}{Australia Telescope National Facility, P.O.Box 76, Epping NSW 1710, Australia} 
\altaffiltext{3}{Harvard-Smithsonian Center for Astrophysics, 60 Garden Street, Cambridge, MA 02138, USA}
\altaffiltext{4}{National Taiwan University, Taipei, Taiwan 106}

\email{mchen@asiaa.sinica.edu.tw}


\begin{abstract}

The Y. T. Lee Array for Microwave Background (AMiBA) has reported the first science results on the detection of galaxy clusters via the Sunyaev Zel'dovich effect.  The science objectives required small reflectors in order to sample large scale structures (20') while interferometry provided modest resolutions (2').  With these constraints, we designed for the best sensitivity by utilizing the maximum possible continuum bandwidth matched to the atmospheric window at $86-102$~GHz, with dual polarizations.  A novel wide-band analog correlator was designed that is easily expandable for more interferometer elements.  MMIC technology was used throughout as much as possible in order to miniaturize the components and to enhance mass production.  These designs will find application in other upcoming astronomy projects.  AMiBA is now in operations since 2006, and we are in the process to expand the array from 7 to 13 elements.

\end{abstract}




\section{Introduction}

AMiBA \citep{kyl02,pho09} is a sensitive interferometer operating at 3~mm wavelength to study the spatial structure of the Cosmic Microwave Background (CMB). The goal is to complement the existing and planned projects in cosmology, such as the Degree Angular Scale Interferometer (DASI) \citep{hal02}, the Cosmic Background Imager (CBI) \citep{pad02}, and the Wilkinson Microwave Anisotropy Probe (WMAP) \citep{ben03}. In particular, AMiBA is imaging clusters via the Sunyaev-Zel'dovich effect (SZE) \citep{sun70, sun72} for the first time at 3~mm wavelength. Our instrumentation is realized due to the maturity of low-noise, wide-bandwidth microwave components in the W band ($75-110$~GHz), as demonstrated by \citet{eri99} and \citet{pos00}. The choice of instrument has been fostered by our technical expertise and interest that grew out of the previous collaboration on the Submillimeter Array project \citep{pho04}.

Radio interferometry is inherently a signal-differencing instrument. It has the advantage in stability and the inherent spatial filtering function, which are suitable for observing CMB and SZ targets with very low systematic error. It has the capability to reject common mode noise usually present in a multi-element array detector system, and to reach intrinsically fainter sensitivities. These technical characteristics attract a number of cosmology projects utilizing interferometry technique \citep{car02}. AMiBA follows the lead of these earlier projects but we exploit the 3~mm band. 

The performance of heterodyne receivers in noise temperature is already near its theoretical limit at 3~mm. To further improve sensitivity, we have to increase the number of the antennas and/or broaden the instantaneous bandwidth of observation. Thus, we opted to build a scalable system with as broad a bandwidth as possible. We had specified an IF of $2-18$~GHz based on commercial availability.  A larger bandwidth would have been much more expensive in construction cost. We matched this IF window to the atmospheric window at $86-102$GHz.

The AMiBA science goals and design philosophy are presented in \citet{pho09}. The telescope and its operational performance are described in \citet{koc09}. \citet{lin09} gives further details on the observational performance of AMiBA. The early science results, carried out with the $7 \times 0.6$~meter antennas, have been reported in a series of companion papers \citep{nis09, wu09, ume09}. The Array was dedicated with its compact 7-element configuration in 2006, and the instrument has been constantly operating ever since. We are now replacing the reflectors with their 1.2-meter counterparts. We are also expanding the array from 7 to 13 elements. The following section gives an overview on the components of the AMiBA instrumentation.

\section{INSTRUMENTATION}

\begin{figure}[t]
\begin{center}
\plotone{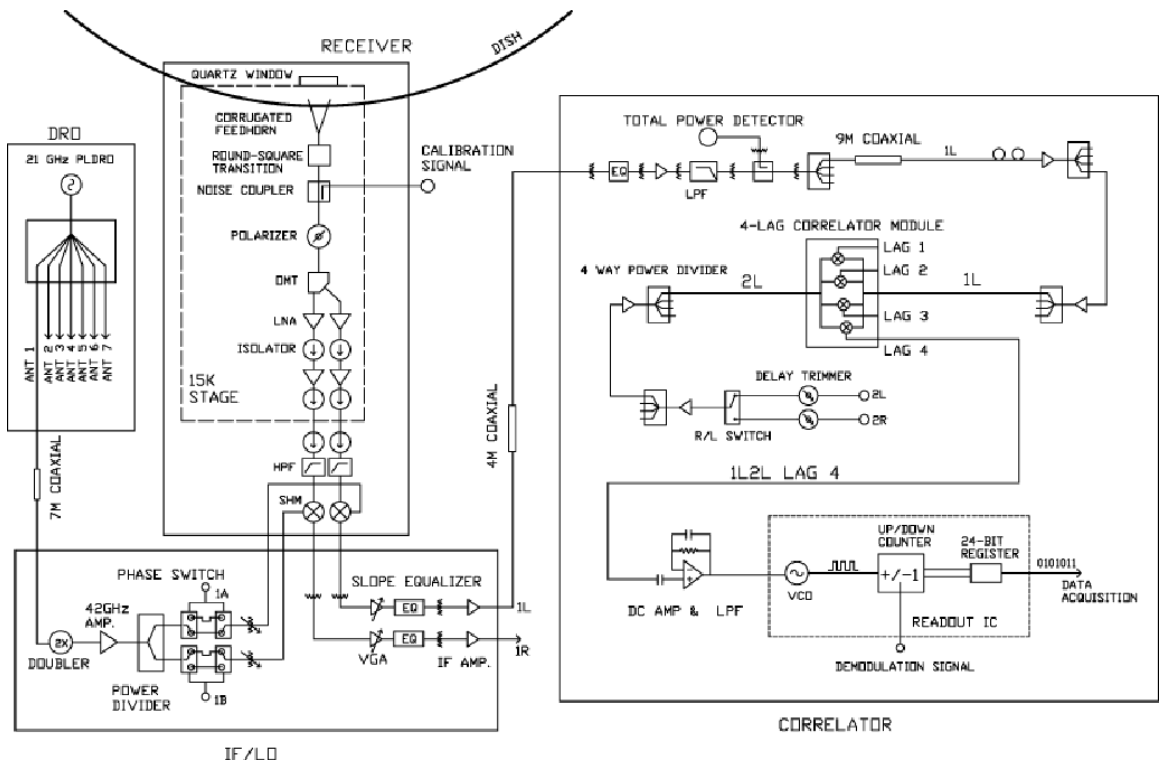}

%%\special{psfile=amiba_block_diagram_2008_c.pdf angle=0 hscale=120 vscale=120 hoffset=-115 voffset=-620}
%%\vskip4.0in

\caption{System block diagram for the AMiBA receiver and correlator configuration.}
\label{Block}
\end{center}
\end{figure}

We designed, tested, and prototyped our components to optimize broadband sensitivity \citep{mtc02, ctl04}. The current instrument specifications are listed in Table~\ref{character}, and Figure~\ref{Block} shows the block diagram of the overall system. We describe below the various features which optimized AMiBA sensitivity, and which can be further enhanced in the future for other applications.   

{\bf Receiver Optics.}
The receiver optics is composed of a Cassegrain reflector coupled to a corrugated scalar feed. The feed has a semi-flare angle of 14 degrees to ensure an approximate constant beam width over the receiver band \citep{pad02} \citep{cla84}. Based on a quasi-optics method described in \citet{gol98}, the feed is designed to illuminate the reflector with an edge taper of 10.5 dB. The beam waist produced from this feed is located at the vertex of the Cassegrain dish.  The advantage for such arrangement is that we can outfit the receiver with reflectors of different apertures. For the physical layout, this feed-horn is similar in design to those used in the Sub-Millimeter Array receivers \citep{zha93} for achieving wide-band single mode operation, and wide-band, low return loss simultaneously.

The antenna is a small-size, Cassegrain reflector made of Carbon-Fiber-Reinforced-Plastic (CFRP). We have two sizes of reflectors, 0.6~meter and 1.2~meter, for different scientific targets. Having several antennae in a close-packed configuration on the platform can cause cross-talk problems. Thus, our reflectors have baffles attached with a height of about 30\% of the diameter in order to suppress crosstalk and ground pickup.  The edges of the baffles are slightly rolled in order to suppress diffraction effects.  The design and fabrication of these reflectors in Taiwan established a team which has gone on to produce nutating subreflectors for the Altacama Large Millimeter/submillimeter Array (ALMA) project under construction in Chile.    

The reflector's surface is coated with aluminum. Aiming at minimizing a possible emission from the underlying material and maximizing the reflection of the incoming signal, an aluminum layer attenuation of less than 1\% is targeted. A 5 times skin depth leads to a $e^{-5}$=0.67\% attenuation, which translates into about a 1.4~$\mu$m coating layer at the relevant frequency. By asking for at least 2.8~$\mu$m coating we are ensuring  a very low attenuation.  

After assembled, the geometry of the primary and the secondary reflector is verified directly on a Zeiss Prismo 10 metrology machine. For the sizes of the AMiBA reflector, the measuring accuracy is better than 5~$\mu$m. %We have requires a surface specification of 50~$\mu$m RMS to ensure at least of 95\% gain.
We have specified a surface accuracy of 50~$\mu$m RMS to ensure a gain efficiency of at least 95\%. The subreflector positioning requirements are based on the Ruze formulae \citep{ruz66}. An axial and lateral secondary defocus of  0.1~$\lambda$ and 0.45~$\lambda$, respectively, keep the gain loss at less than 1\%. ($\approx$  3~mm at 90~GHz.)  Similarly, a feed horn positioning within 1~$\lambda$ gives a 99\% gain. 

{\bf Dual Polarization Front-end.}
The receiver front-end contains the scalar feed followed by a noise coupler, a phase shifter and an orthomode transducer (OMT) for polarization separation \citep{wol02}. A short section of 90-degree phase shifter is connected to the OMT to perform linear-to-circular polarization transformation. Currently this phase shifter is replaced with a simple waveguide transition, thus we are observing in linear polarization detections. These OMTs show typical insertion losses of less than 0.2~dB, return loss around 20~dB, and more than 40~dB isolation over the $75-110$~GHz band. Similar OMTs have been used in the WMAP and ALMA \citep{cla05}.  The availability of dual polarizations increases our sensitivity by a factor of $\sqrt{2}$, and will enable studies of polarization in the future.  

After the OMT, each polarized signal then passes through its own chain of amplifiers, isolators, and a high-pass filter before reaching a sub-harmonic mixer. Two cascade amplifier modules and two wideband WR-10 cryogenic isolators form the RF amplifier chain. Each amplifier module contains a four-stage, InP high electron mobility transistor (HEMT), which is fabricated as a monolithic millimeter-wave integrated circuit (MMIC) \citep{wei99}. Measured at 20~K cryogenic temperature, its noise temperature is typically 40-50K, with associated gain higher than 22~dB over $80-105$~GHz. The same LNA chip has been used in several other astronomical experiments; e.g. the Cosmic Anisotropy Polarization Mapper (CAPMAP) of Princeton University, the Sunyaev-Zeldovich Array (SZA) of the University of Chicago, and the SEQUOIA from the University of Massachusetts \citep{eri99}.

The filter before the mixer is designed with the cutoff frequency at 84.5~GHz \citep{liu03}. It enables an upper-sideband conversion for the following mixer. Developed in-house \citep{hwa02a, hwa02b}, a sub-harmonically-pumped (SHP) diode mixer module mixes the $86 - 102$~GHz RF signal with the 42~GHz LO and generates an IF of $2 - 18$~GHz. This mixer has achieved a flat response in the conversion loss of around 12~dB at an optimal LO power level of 7~dBm.  The development of MMIC technology in Taiwan enhances the potential to mass produce large format array detectors.  This can drive the future of multi-pixel heterodyne interferometry.  The HEMT technology is also being developed for ALMA Band 1 receivers.  

While the entire front-end is enclosed in a vacuum chamber, only the components prior to the high-pass filter are cooled by the CTI-22 coldhead. This is to minimize the thermal loading on the coldhead, and to achieve a relatively simple mechanical design for the cryogenic environment. We used a z-cut quartz with anti-reflection coating as the vacuum window\footnote{The windows are manufactured by QMC Instrument Ltd.}. This coated quartz window comes in with 3~mm thickness, and 35~mm in diameter.  Its power transmission in the band of our interest has been measured using a vector network analyzer.  With a time-gating technique \citep{edi00}, the signal transmission is measured to be better than 97\% in $80-105$~GHz. 

\begin{figure}
%%\begin{center}
%\plotone{receiver_photo}
\centering
\includegraphics[totalheight=8in]{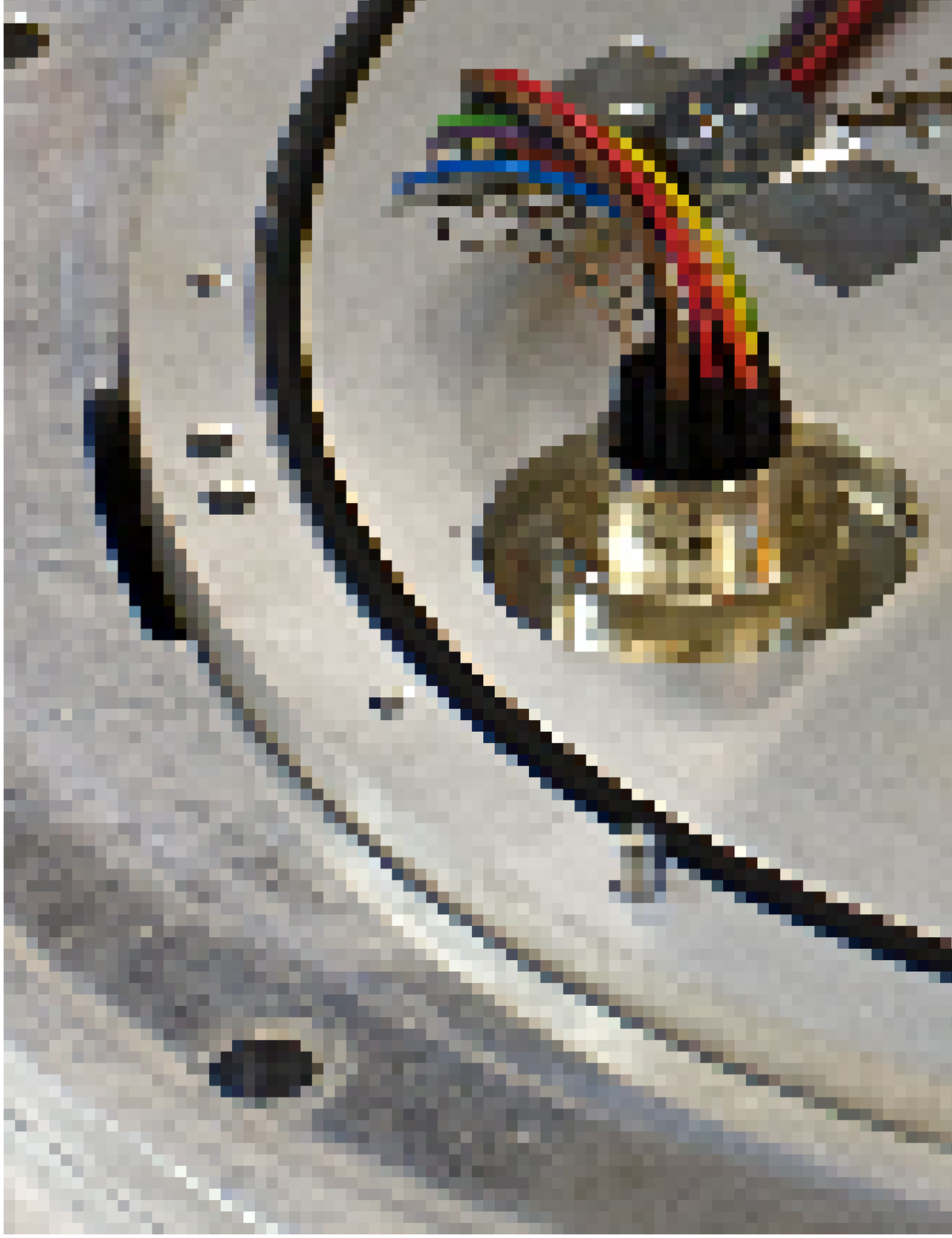}  %original receiver_photo
\caption{AMiBA receiver with vacuum jacket removed.}
\label{receiver}
%%\end{center}
\end{figure}

After down-conversion by the SHP, the IF signals are sent out of the cryostat and fed into the IF/LO module for further signal amplification. 
%The difference in LO paths for phase switching creates an unexpected power variation for pumping the mixers.
LO phase switching is accomplished by using a pair of PIN switches and delay lines. This created an unexpected power variation for mixer LO drive. Although small, the subsequent IF levels become imbalanced and produce a systematic DC offset from the correlator output. The LO is intentionally adjusted to over-pump the mixer to reduce this effect.  The development of broadband LNA and SHP are key to reducing component costs, miniaturization, and lowering the weight of the receiver packages. Figure~\ref{receiver} show the AMiBA front-end with the vacuum jacket removed.

{\bf Signal Distribution and Correlation}

The instantaneous bandwidth of 16~GHz in IF presents a major technical difficulty for the signal processing after the receiver front-end. It is a great technical challenge to effectively distribute these multi-octave signals through an interconnected network. 
%Physically, each of the IF signal paths is designed with two 4-way power dividers and a total of 10-meter, semi-rigid coaxial cable between the SHP mixer and the correlator. 
Physically, each of the IF signal paths is designed with two 4-way power dividers in cascade and a total of 13-meter, semi-rigid coaxial cable between the SHP mixer and the correlator. To compensate for the signal transmission loss and the incurred bandwidth slope, the signal distribution network encompasses a series of power amplifications, power dividing, filtering, and slope correction. Active temperature compensation mechanisms are built into all the enclosed electronic modules. In all, although there is more than 150~dB active power gain in a single path, the IF network only provides a net gain of 36~dB, with a noise figure of about 10~dB between the SHP mixer and the correlator

Following on the development in broadband, lag-correlator by \citet{har01}, we designed and built a 4-lag, analog correlator to achieve the unprecedented correlation bandwidth. The correlator module with four different delays yields complex visibility data in 2 bands. Each lag spacing is 25 pico-seconds to sample signals up to 20~GHz of bandwidth. The decision on the number of the lags was another compromise between the cost and the performance. A doubly balanced diode mixer is chosen as the correlation multipliers to cope with broadband correlation. The passive multipliers avoid problems of 1/f noise, and the circuitry complications associated with active multipliers. The issue of high input impedance associated with non-biased mixer is mitigated with the low barrier diodes chosen for the multiplier design. The diode mixers are usually broadband devices. It is the matching circuitry between the diodes and the input IF signals that dominates the effective bandwidth of the AMiBA correlator.

The correlated outputs are amplified and then digitized by custom-designed readout ICs. Inside the readout IC, voltage-control oscillators (VCOs) implement the voltage to frequency conversion at 8~MHz, and the following up/down counters integrate the clock signal for a short period of time. Each of the IF signal is phased-switched at the down-conversion following an exclusive set of Walsh time sequences. After the signal correlation, the demodulation signal is applied to the up/down counter. The combination of the phase switching and the demodulation process can reject any false signals that do not have the corresponding signature as the demodulation signal.

\section{SYSTEM PERFORMANCE}  

\begin{figure}
\begin{center}
\plotone{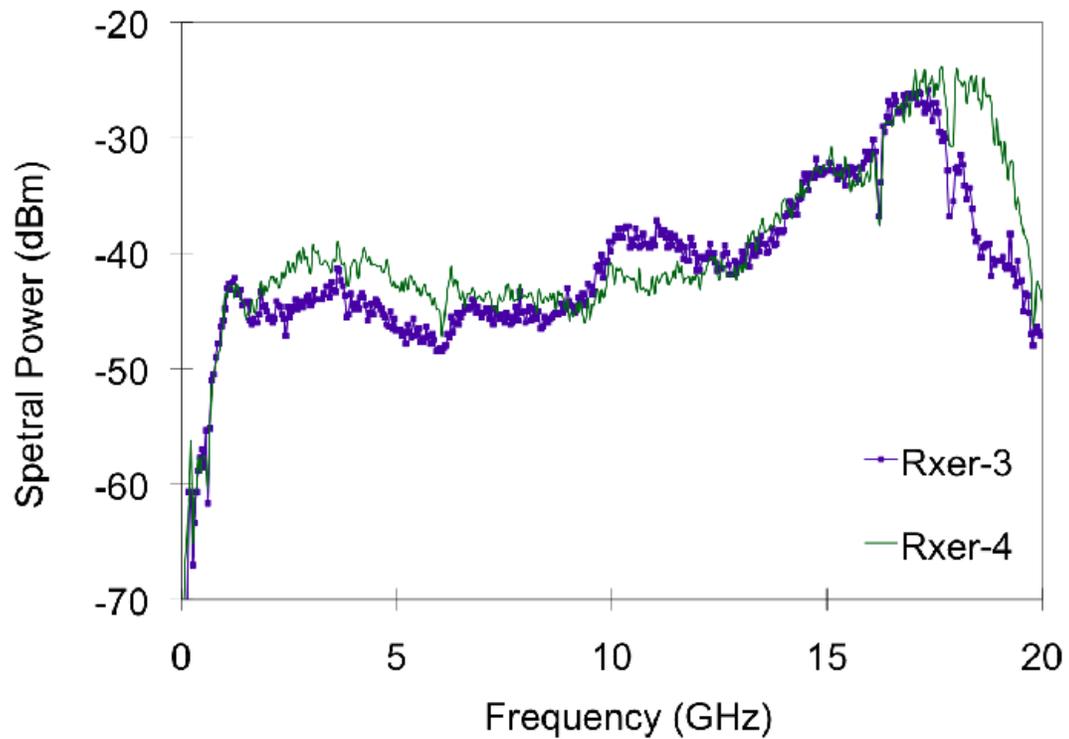} 
\caption{IF Power spectra of the AMiBA receivers No.3 and No.4.}     
\label{spectrum}
\end{center}
\end{figure}

\begin{figure}
\begin{center}

\plotone{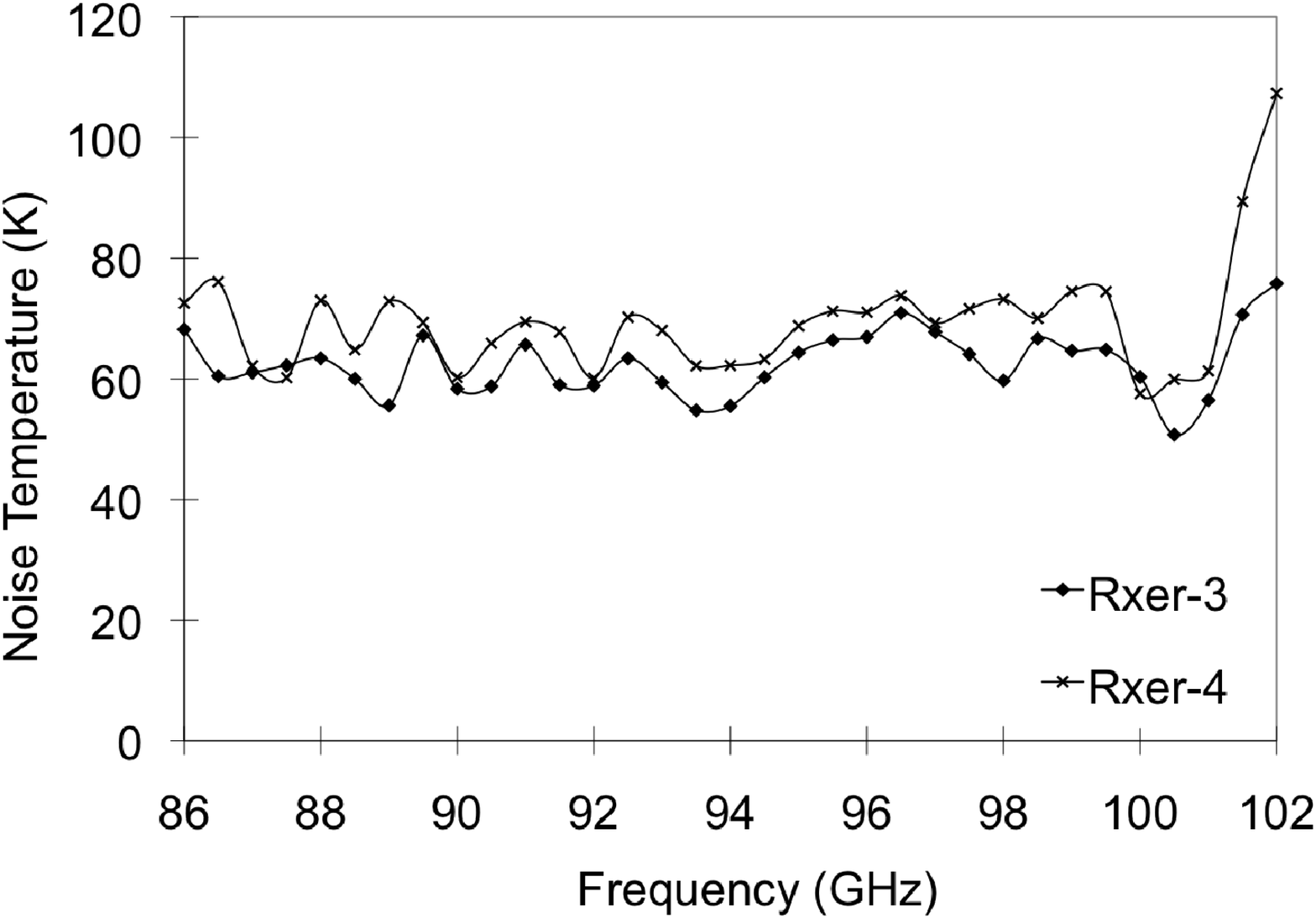}

%\special{psfile=LLRR-allpha.png angle=0 hscale=6 vscale=6 hoffset=-115 voffset=-620}

%\vskip4.0in

\caption{Receiver noise temperatures over the AMiBA RF band on two selected receivers; No.3 and N0.4. The markers represent the data points.}
\label{noise}
\end{center}
\end{figure}

The receiver pass-band typically has considerable larger gain toward high frequency. This is mainly from the gain profile of the %LNAs, 
first slope equalizers in the IF chain, and it actually compensates for the gain slope caused by the long cables in the distribution network. Figure~\ref{spectrum} shows the output spectra on some selected AMiBA receivers. The receiver noise temperature is measured using a conventional Y-factor method at room temperature with liquid-nitrogen-bath absorber as a cold load. Measured after the first IF amplifier, all the receivers show consistent noise performance with values ranging in $55-75$~K, and with slight degradation near the high end of the band. Figure~\ref{noise} plots the noise temperatures over the RF band on two of the selected AMiBA receivers. The LNA typically contributes about 50~K in the overall receiver noise temperature. The rest of the noise contribution is from the passive waveguide losses in front of the first LNA. This non-trivial noise contribution is a compromise from our intention for calibration and for detecting circular polarizations. 

The inputs to the AMiBA correlators is dominantly the uncorrelated noises from the front-ends amplified by the system gain before the correlators. The internal noise of the AMiBA correlators is small compared with these inputs and can be ignored in the system performance consideration. The cross-correlation of these noises in each baseline produces the low frequency fluctuations at the correlator outputs and limits the correlator sensitivity. 
%On the other hand, when there is a common signal present in the baseline, the correlator produces a correlated signals at 1~KHz, at which the signals are phase-modulated.  
On the other hand, when there is a common signal present in the baseline, the correlator produces a correlated signal at 700~Hz, which is the phase modulation rate. The ratio of the root-mean-squared amplitudes of this 700~Hz signal over the low-frequency fluctuations is defined as the signal-to-noise ratio (SNR) for our detection system. Our effort is to tune the system to achieve the best SNR with reliable performance. 

\begin{figure}
\begin{center}
\plotone{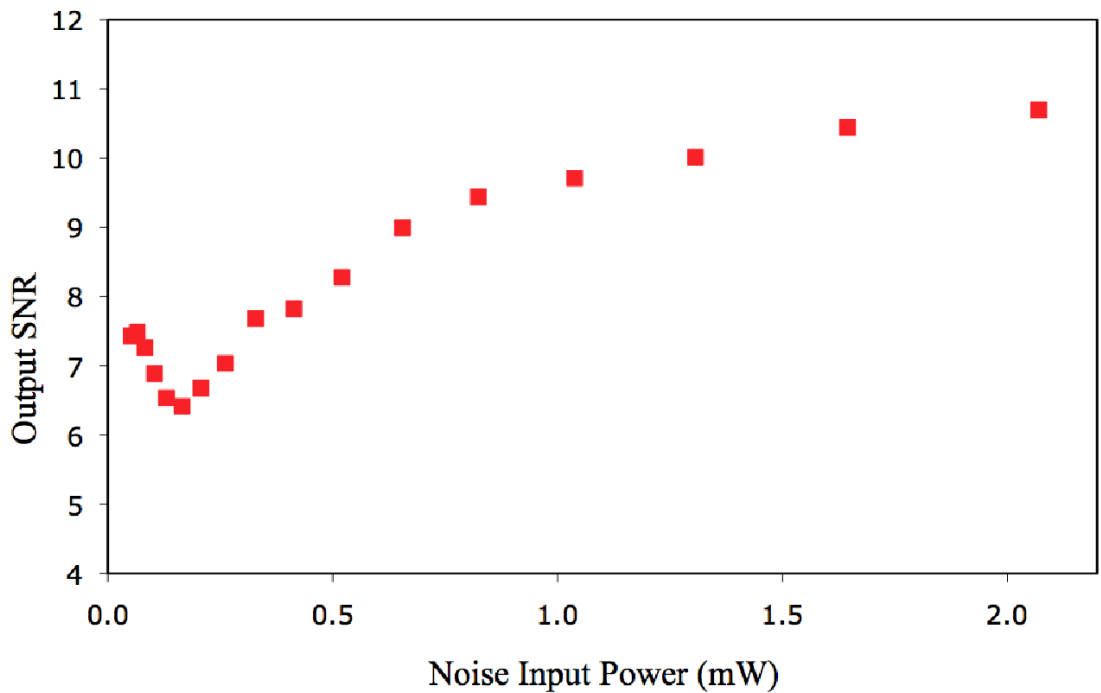} 
\caption{A typical output signal-to-noise ratio of the AMiBA correlator versus balanced input power. This measurement is for determining the nominal input power level to endure an optimal correlator operation.}
\label{SNR}
\end{center}
\end{figure}

\begin{figure}
\begin{center}
\plotone{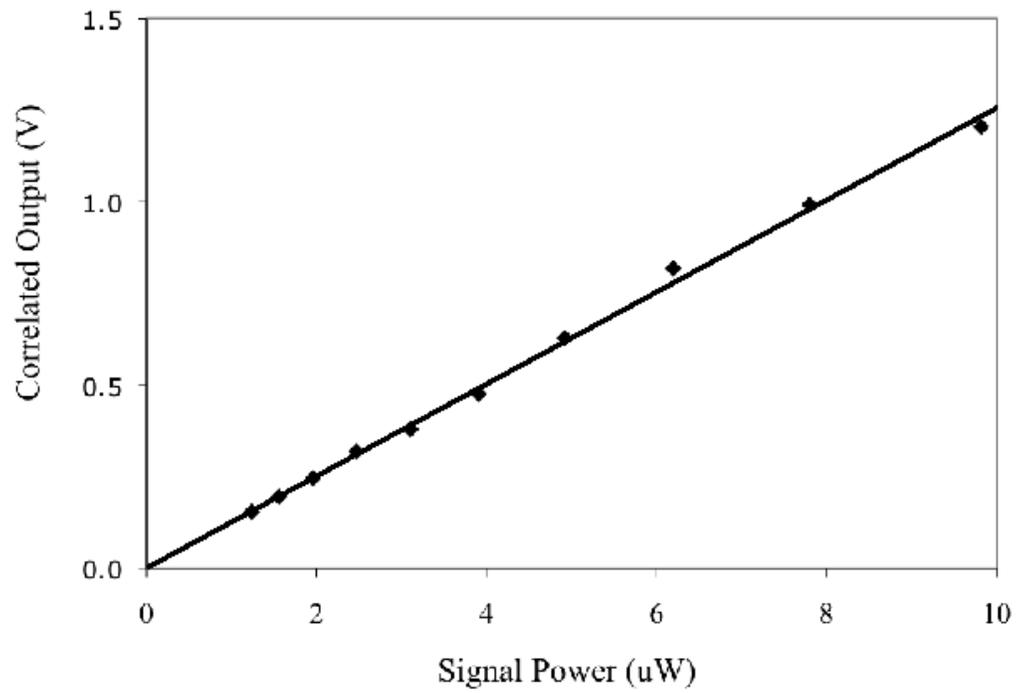}
\caption{AMiBA correlator linearity test. While driven by large, uncorrelated noise at 1 mW, the correlator outputs linearly with its input correlated signals. The line represents the best linear fit through the data points.}
\label{signal}
\end{center}
\end{figure}

A group of randomly selected correlator modules are tested in the laboratory to determine the optimal input powers to the correlator for the best output SNR. The test is to measure the correlator's output SNR versus %the same level of the
input power applied to the correlator. To simulate the observation, the input for the test consisted of a large uncorrelated noises mixed with a small common signal. The input SNR is kept at a nominal value of $-20$~dB throughout the test. All the correlators tested show very similar characteristics. One set of the test results are depicted in Figure~\ref{SNR}. Within the tested power range, the output SNR has a relatively steep minimum around 0.2~mW, and then increases and approaches a saturated value at the high input beyond 2~mW. The increase in the output SNR at high driving power is mostly due to the increase in the output signal level, while the RMS level of the low-frequency fluctuations stays almost the same beyond 0.5~mW. We have concluded that the AMiBA correlator modules deliver better output SNR with high level of uncorrelated driving power. In a following test, we measured the signal output linearity particularly at high driving power. As depicted in Figure~\ref{signal}, the output signal level remains in very good linearity with the balanced input signal, while the correlator is driven by high noise power at 1~mW. These results suggest that we should set up a high-gain power system before the correlators.

In reality, it is very difficult to maintain balanced inputs to the correlators, especially in a wide band system \citep{pad94}. One must consider the imbalance of the IF power levels due to the LO switching and the different gains of the signal path. More seriously, there is the passband mismatch between the IF signals \citep{tho82}. We have measured the correlator's output SNR versus the imbalanced input power levels. 
%In this test, the input SNR is still fixed at $-20$~dB, while the input power levels to the correlator are varied in increment of 3~dB, or by a factor of 2. The output SNR from these measurements degrades from its nominal value with balanced power inputs. However, The degradation sharply increases when both the input power is beyond roughly 2~mW, or below  0.25~mW. Within this range, the SNR reduction remains in about $10-15$\%. 
In these measurements, the input SNR is still fixed at $-20$~dB, while the pair of inputs to the correlator differ in power by 3~dB, or by a factor of 2, with the power ranging from -6~dBm to 3~dBm. The output SNR from these measurements degrades from its nominal value with balanced power inputs; however, the reduction in SNR remains in an acceptable level of $10-15$\%. 

By observing Jupiter, we have verified that better SNR could be achieved with high input power level into the correlators. The IF power imbalances are carefully controlled with the power alignment in each receiver channel, with the active temperature stabilization in the electronics, and with the adjustment of the variable gain amplifiers in the IF signal paths. The power imbalances in our system are controlled to be within 1.5~dB during a typical observation. Thus, practically, we have set the input power levels to the correlators at a nominally 0.5~mW, or -3~dBm. 

Because of the wide bandwidth and non-monotonic responses of microwave components, it is necessary to calibrate the pass-band properties for each baseline. This information is also essential for lag-to-visibility transformation. The correlator pass-band is recovered by spatially sweeping a broadband noise source across a pair of the receivers. The cross-correlation output is measured against the relative time delay between the receivers, and is Fourier-transformed to yield the spectral response of the passband. Dominated by the broadband matching circuit, the spectral structure typically shows a major gain dip in $6-12$~GHz. At some extreme frequencies, the gains are off the peak value by more than 10~dB. The large fluctuations in the correlator passbands result in a narrower effective bandwidth.  Further details in this part of work will be addressed in \citet{lin09}. A correlator module with a flat spectral response across the IF band would yield a wider effective bandwidth, and thus increase the AMiBA instrument sensitivity. Alternatively, increasing the number of the lags would have had a similar benefit to our system.

%\begin{figure}
%\begin{center}
%\plotone{XX-allamp} 
%\plotone{XX-allpha}

%\caption{Cross-correlation spectral responses from the relative time-delay measurements of each pairs of the receivers.}     
%\label{corr-band}
%\end{center}
%\end{figure}

\section{CONCLUSION}

In pursuing high sensitivity and low systematics, we have built a radio interferometer to observe the CMB and the SZE at 3~mm wavelength. The number of elements and the instantaneous bandwidth are the two critical parameters influencing the system performance of AMiBA. The sensitivity of the interferometer is roughly proportional to the number of array elements. The MMIC technology addresses component miniaturization, production capability, performance uniformity, and the state-of-the-art sensitivity. Its rapid progress in millimeter wavelength has already provided realistic ways to build compact receivers with mass production capacity, such as in the Q/U Imaging Experiment (QUIET) \citep{sam08}. The advantage of MMIC would benefit especially large-scale projects, such as the ALMA Band-1 and Band-2, as well as the Square Kilometer Array.

Large instantaneous bandwidth increases the sensitivity with its square-root of the increment. For continuum sensitivity, the lag-correlator has the advantage, among other correlation techniques, to achieve very wide bandwidth operation. The major challenge is in the availability of broadband devices and circuitries to handle multi-octave signals.

At millimeter wavelengths, heterodyne detection with a broad correlation bandwidth as in AMiBA would be competitive with bolometric techniques in terms of system noise and sensitivity. In addition to cosmology, a broadband, lag-correlators, with only coarse to moderate spectral resolution, may be suitable for red-shift surveys of extra-galactic objects at submillimeter wavelengths, or in the tera-hertz regime. By increasing the number of lags, it may also be practical for Galactic observations.

This is an overview on the detection system for the 7-element AMiBA array. The system has been under routine operation during the past two years at the site in the Mauna Loa Observatory, Hawaii. We are in the process to expand this array to 13 elements with 1.2-meter reflectors. 

%The initial science results are published together with this issue. The further details on the observational performance of AMiBA can be found in \citet{lin09}, also in this issue. 

{\bf Acknowledgments.} 

We thank the Ministry of Education, the National Science Council, and the Academia Sinica for their support of this project. We thank the Smithsonian Astrophysical Observatory for hosting the AMiBA project staff at the SMA Hilo Base Facility.  We thank the NOAA for locating the AMiBA project with their site on Mauna Loa.  We thank the Hawaiian people for allowing astronomers to work on their mountains in order to study the Universe.

\clearpage

\begin{deluxetable}{ll}
  \tablecaption{AMiBA Detection System Characteristics\label{character}}
  \tablehead{
    \colhead{Components}        & \colhead{Specifications}
  }

  \startdata
  Cassegrain Reflector, f/2.0   & 0.6-meter and 1.2-meter Interchangeable \\
  Receiver Front-end            & MMIC HEMT LNA \\
                                & Dual-polarization \\
                                & Cooled to 15~K \\
  RF			        & $86 - 102$~GHz \\
  Receiver Noise Temperature    & $65 \pm 10$~K\\
  Down-conversion               & Sub-harmonic mixer\\
  LO Frequency                  & 42~GHz with phase-switching\\
  LO Source                     & Dielectric Resonator Oscillator at 21~GHz\\
  IF                            & $2 - 18$~GHz\\
                                & Variable Gain Control\\
  Correlator                    & Analog correlation\\
   		                & 4 lags\\
                                & Diode multiplier\\
  \enddata
\end{deluxetable}

\clearpage

\clearpage

\clearpage

\clearpage

\clearpage

\clearpage

\end{document}